# AlGaN Channel Field Effect Transistors with Graded Heterostructure Ohmic Contacts

Sanyam Bajaj[1,a)], Fatih Akyol[1], Sriram Krishnamoorthy[1], Yuewei Zhang[1] and Siddharth Rajan[1]

[1]*Department of Electrical & Computer Engineering, The Ohio State University, Columbus, OH 43210*

**Abstract:** We report on ultra-wide bandgap (UWBG) $Al_{0.75}Ga_{0.25}N$ channel metal-insulator-semiconductor field-effect transistors (MISFET) with heterostructure engineered low-resistance ohmic contacts. The low intrinsic electron affinity of AlN (0.6 eV) leads to large Schottky barriers at metal-AlGaN interface, resulting in highly resistive ohmic contacts. In this work, we use reverse compositional graded $n^{++}$ AlGaN contact layer to achieve upward electron affinity grading, leading to a low specific contact resistance ($\rho_{sp}$) of $1.9 \times 10^{-6}$ $\Omega.cm^2$ to n-$Al_{0.75}Ga_{0.25}N$ channels (bandgap ~ 5.3 eV) with non-alloyed contacts. We also demonstrate UWBG $Al_{0.75}Ga_{0.25}N$ channel MISFET device operation employing compositional graded $n^{++}$ ohmic contact layer and 20 nm atomic layer deposited $Al_2O_3$ as the gate-dielectric.

a) Author to whom correspondence should be addressed. Electronic mail: bajaj.10@osu.edu Tel: +1-614-688-8458

III-Nitrides offer a broad spectrum of bandgaps from 0.6 eV (InN) to 6.2 eV (AlN), and have various device applications ranging from optoelectronics to RF amplification and power switching applications. Due to the superior intrinsic material properties of AlN, wide bandgap AlGaN with high Al composition could enable next-generation RF power amplifiers and switches. The large bandgap of AlN results in a theoretical breakdown field of 12-16 MV/cm, which is almost 4-5 times higher than that of GaN[1]. Furthermore, Monte-Carlo calculations have estimated electron saturation velocity in AlGaN channels to be comparable to that in GaN channel[2,3], making AlGaN a promising material for high power and high frequency applications. For power switching application, previous analytical calculation suggested that replacing conventional GaN channel with high Al composition AlGaN channel in high electron mobility transistors (HEMTs) could result in reduced switching losses and enhanced normally-off operation[4]. Although previous experimental reports on AlGaN channel HEMTs have showed encouraging on-state and off-state device characteristics[5-9], a critical challenge preventing further advancement in ultra-wide bandgap (UWBG) AlGaN-based devices was high-resistance ohmic contacts. Several studies have investigated alloyed ohmic contacts to n-AlGaN channels. While alloys using metals such as Titanium, Vanadium and Zirconium[10-15] were reported with low specific contact resistance values below $10^{-5}$ $\Omega.cm^2$ on n-AlGaN channels with Al alloy compositions up to 66%, they are challenging to reproduce reliably due to extremely high temperature processes and typically show non-uniformity in current-voltage characteristics. Alternate heterostructure engineering approaches are therefore needed to realize low contact resistance to large bandgap AlGaN.

A low-resistance ohmic contact is formed by reducing the potential barrier between a metal and semiconductor. The ideal n-type ohmic contact would have a zero or low Schottky barrier height at the metal-semiconductor interface, which can be achieved by matching the semiconductor electron affinity and metal work function. However, the intrinsic low electron affinity of AlN (0.6 eV) leads to larger metal-AlGaN Schottky barriers, resulting in a poor tunneling probability for electrons (probability$\sim e^{\sqrt{\varphi_b}.W}$, where $\varphi_b$ is the barrier height and W is the tunneling width), and therefore highly resistive ohmics. In this work, we take heterostructure engineering approach in which the Al alloy composition in the AlGaN channel is graded from wider bandgap to narrower bandgap under the ohmic contacts, hence grading up electron affinity and presenting a higher electron affinity at the metal-semiconductor interface (GaN electron affinity, $\chi_{GaN}$=4.1 eV). AlGaN layers with compositional grading from GaN to AlGaN have been studied extensively, and shown to induce bulk three-dimensional electron distributions due to positive polarization (spontaneous+piezoelectric) charge[16-19]. The polarization-induced fixed charge, $\rho_\pi$, $\rho_\pi = -\nabla \cdot \boldsymbol{P}$, where $\boldsymbol{P}$ is the sum of spontaneous and piezoelectric polarization in AlGaN alloy. In case of layers with reverse Al compositional grading from wider to narrower bandgap AlGaN, a negative polarization charge is formed, causing a positive curvature in the energy band profile, and thereby creating a barrier to electron flow. To ensure that the conduction band stays flat, it is necessary to compensate for the negative polarization charge using donors. This is shown in the charge diagram in Figure 1, where donors in the graded region compensate for the negative charges, leading to an effectively n-type region. The entire energy band gap offset is supported in the valence band, as shown in the energy band diagram, and there are no heterojunction or electrostatic barriers to transport between the channel and the surface of the semiconductor.

AlGaN structures were grown on Al-face AlN on Sapphire templates[20] using plasma-assisted molecular beam epitaxy (PAMBE) as shown in Fig. 2(a). The 100 nm n-Al$_{0.75}$Ga$_{0.25}$N channel (Si=3x10$^{19}$ cm$^{-3}$) was grown at 720°C on a 30 nm undoped Al$_{0.75}$Ga$_{0.25}$N buffer layer, followed by 50 nm n$^{++}$ contact layer formed by linearly grading down the Al content from 75% to ~0% (actual 6%). Figure 3(a) shows the X-ray diffraction scan measured with AlN as the reference to confirm the Al composition and thickness of the epi layers. Atomic force microscopy (AFM) on the as-grown surface, as shown in Fig. 3(b), indicated a fairly smooth surface morphology with rms roughness of 1.1 nm and a step-flow growth mechanism. Non-alloyed Ti/Al/Ni/Au ohmics were evaporated on as grown structures, followed by device mesa isolation of 200 nm using $Cl_2$-based inductive plasma etching. To achieve active $Al_{0.75}Ga_{0.25}N$ channel and characterize the ohmic contact, the compositional graded contact layer was recessed between the source and the drain of the device using low-power (6 Watts) $Cl_2$-based inductive plasma etching.

Figure 2 (b) shows the energy-band diagram in the contact region (as-grown), simulated using 1D Schrodinger-Poisson simulator[21]. It can be seen that the conduction band ($E_C$) profile under the contacts does not have any abrupt or electrostatic barriers to block the flow of electrons. Hall measurement on as-grown structures with both channel and contact layers gave a sheet resistance ($R_{SH}$) of 160 Ω/□, and net effective mobility of 35 $cm^2$/Vs. Figure 4 (a) illustrates the transport and resistance components in the as-grown structures. Transfer length measurement (TLM), shown in Fig. 4 (b), gave an $R_{C1}$ of 0.15 Ω.mm, which is the contact resistance at the metal-$n^{++}$ $Al_{0.06}Ga_{0.94}N$ junction. To test the contact to $Al_{0.75}Ga_{0.25}N$ channel, the measurements were repeated after recessing the graded contact layer and leaving 90 nm thick channel. Hall measurement on the recessed structure gave a sheet resistance ($R_{SH}$) of 726 Ω/□, and a low channel mobility of 16 $cm^2$/Vs, feasibly limited by impurity scattering effect due to high Si donor concentration and native defects in the channel layer. A high Si concentration was used in the channels since samples with lower doping below $10^{19}$ $cm^{-3}$ were insulating, suggesting an acceptor-like compensating defect with concentration of the same order. While secondary ion mass spectrometry (SIMS) revealed much lower Carbon and Oxygen background concentrations below $3 \times 10^{17}$ $cm^{-3}$, the source of compensation is not known at this point and further research is being carried out to investigate and improve the channel mobility. To measure the effective contact resistance to n-Al$_{0.75}$Ga$_{0.25}$N channel, TLM was performed and plotted against varying recess spacings, as shown in Figure 5. The effective contact resistance in the recessed device geometry is given by the summation of R$_{C1}$ (Fig. 4 (b)), and R$_{SH1}$ which consists of the sheet resistance components of the graded contact layer and the channel layer, as illustrated in Fig. 5 (a). TLM gave R$_{SH}$ values of 725 Ω/□, and R$_C$ of 0.32 Ω.mm, as shown in Fig. 5 (b). The specific contact resistance ($\rho_{sp} = R_C^2/R_{SH}$) was extracted to be 1.9x10$^{-6}$ Ω.cm$^2$, which is the lowest value reported to AlGaN with such high bandgap of 5.3 eV using non-alloyed ohmic contacts, and is comparable to typical values achieved on lower band gap GaN channels[22,23].

To demonstrate MISFET device operation, identical structures with thinner (20 nm) n-Al$_{0.75}$Ga$_{0.25}$N channels were grown for better recess control and to achieve a lower channel charge suitable for transistor operation, mainly device pinch-off and high breakdown voltage. The graded contact layer was recessed between source/drain pads to form active channel with thickness ~ 12 nm confirmed using AFM. The device process was concluded with a 20 nm ALD Al$_2$O$_3$ gate-dielectric deposited at a substrate temperature of 300°C, followed by post-deposition anneal at 700°C to minimize the Al$_2$O$_3$/AlGaN interface defect states or hysteresis[24], and deposition of Ni/Au/Ni gate metal. Figure 6 shows the final MISFET device structure and experimental characteristics for gate-length, $L_G$ = 0.7 µm and gate-drain spacing, $L_{GD}$ = 1.1 µm. Fig 6 (b) shows the capacitance-voltage measurement (10 kHz frequency) on a circular (100µm radius) diode structure, with MESFET-like behavior in accumulation region and a pinch-off voltage of -8 V. The measured family output I-V characteristics are shown in Fig. 6 (c), with maximum drain current, $I_{DS\_MAX}$ of ~ 60 mA/mm at $V_{GS}$ = 2 V. The measured transfer characteristics at $V_{DS}$ = 20 V are shown in Fig. 6 (d), with peak transconductance, $g_{m\_peak}$ of 14 mS/mm. The $I_{DS\_MAX}$ and $g_{m\_peak}$ values are largely limited by the low channel mobility, and are expected to improve with enhanced material quality (lower native defect / Si dopant concentrations). Finally, the breakdown performance was evaluated by measuring 3-terminal off-state breakdown at 1 V below pinch-off voltage. At $V_{GS}$ = -9 V for a gate-drain spacing of 1.1 µm (confirmed using scanning electron microscope), a 3-terminal breakdown voltage, $V_{br}$, of 224 V was measured, as shown in Figure 7, in the presence of Fluorinert solution but without any field plates or edge termination techniques. Although the measured breakdown voltage of approximately 200 V/µm is significantly lower than theoretically predicted values, it is nevertheless among the highest for AlGaN channel devices, and is expected to improve with engineering the peak electric field in the channel (eg. field plates) and superior material quality.

In summary, a low specific contact resistance ($\rho_{sp}$) of $1.9 \times 10^{-6}$ $\Omega.cm^2$ was achieved on UWBG n-$Al_{0.75}Ga_{0.25}N$ channels using heterostructure engineered non-alloyed ohmics. Reverse compositional graded n$^{++}$ AlGaN layer eliminated Schottky barrier at the metal-semiconductor interface and resulted in flat conduction-band profile posing no energy barriers to the electron carriers. This approach employed non-alloyed ohmic contacts which have enormous advantage of enabling self-aligned and Au-free device processes. This result could also have applications in a large range of AlGaN-based electronic and photonic devices, and advance the research in the area of high Al composition AlGaN materials and devices. We also demonstrated UWBG $Al_{0.75}Ga_{0.25}N$ channel MIS transistor with heterostructure engineered ohmic contacts and 20 nm $Al_2O_3$ gate-dielectric, with an $I_{DS\_MAX}$ of 60 mA/mm, $g_{m\_peak}$ of 14 mS/mm and $V_{br}$ of 200 V/µm.

Acknowledgement:

The authors thank W. Sun, M.S.H. Sohel, and Prof. Arehart for their help with high voltage measurements. The authors appreciatively acknowledge support from National Science Foundation (ECCS-1408416), Office of Naval Research (Dr. Paul Maki) and Raytheon IDS Microelectronics.

Figure captions:

**Figure 1:** Structure schematic, energy-band diagram and charge diagram for reverse compositional graded n$^{++}$ AlGaN contact layer on UWBG n-AlGaN channel. Negative polarization charge in the reverse graded AlGaN layer is compensated by donors, resulting in flat conduction-band profile under the ohmic contact.

**Figure 2:** (a) Schematic of AlGaN structures used for the study, with 50 nm reverse compositional graded n$^{++}$ contact layer above the 100 nm n-Al$_{0.75}$Ga$_{0.25}$N channel; (b) the associated energy-band diagram under the ohmic contact region (as-grown) showing upward grading of electron affinity towards the surface.

**Figure 3:** (a) Measured X-ray diffraction scan confirming AlGaN compositions and thicknesses; and (b) Atomic force microscopy on as-grown surface showing fairly smooth surface morphology and step-flow growth mechanism.

**Figure 4:** (a) As-grown n-Al$_{0.75}$Ga$_{0.25}$N MESFET structure depicting current flow, contact resistance, $R_{C1}$, and net sheet resistance, $R_{SH}$ in the device; and (b) Transfer length measurement on the structure in (a) with measured values of $R_{C1}$ and $R_{SH}$.

**Figure 5:** (a) Gate-recessed n-Al$_{0.75}$Ga$_{0.25}$N MESFET structure with 90 nm channel layer between the ohmic contacts, depicting current flow, effective contact resistance, $R_C = R_{C1}+R_{SH1}$, and channel sheet resistance, $R_{SH2}$; and (b) Transfer length measurement on the structure in (a) with measured values of $R_C$ and $R_{SH2}$.

**Figure 6:** (a): Structure schematic of the Al$_{0.75}$Ga$_{0.25}$N channel MISFET; (b) measured capacitance-voltage characteristics (10 kHz frequency) on circular diode structures (radius = 100 μm) and the integrated charge density as a function of gate-bias; (c) measured family output I-V characteristics; and (d) measured transfer characteristics at $V_{DS}$ =20 V; device gate-length, $L_G$ =0.7 μm and gate-drain

spacing, $L_{GD}$ =1.1 µm.

**Figure 7:** 3-terminal breakdown voltage, $V_{br}$ = 224 V measured at $V_{GS}$ = -9 V for a gate-drain spacing, $L_{GD}$ = 1.1 µm.

1.

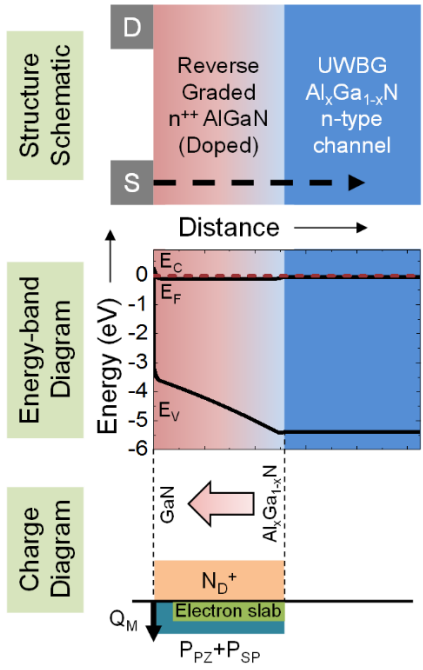

2.

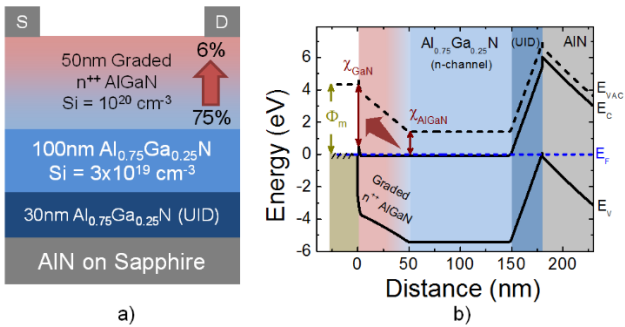

3.

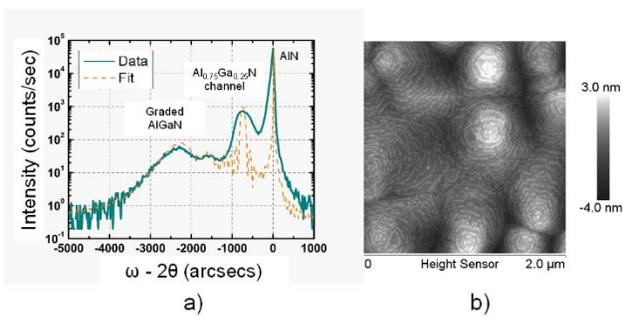

4.

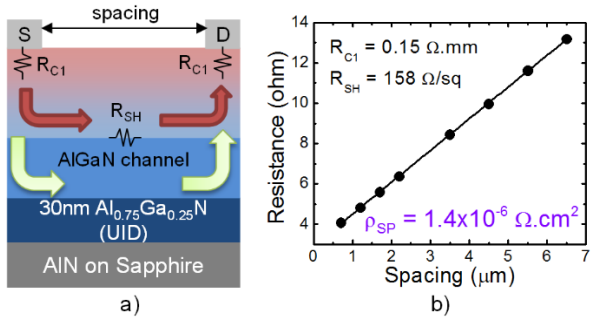

5.

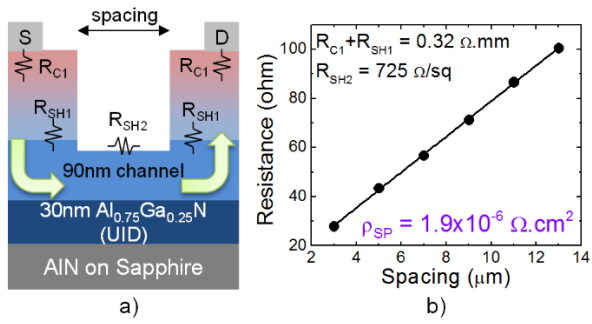

6.

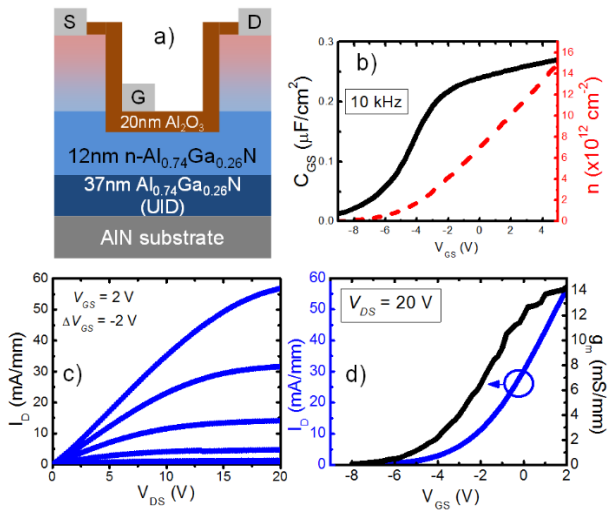

7.

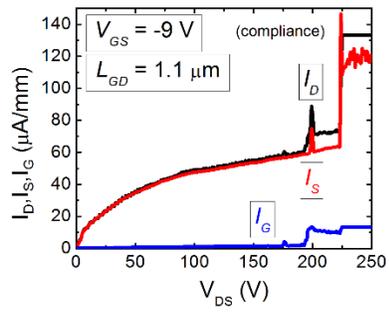